\documentclass[conference]{IEEEtran}
\IEEEoverridecommandlockouts
\usepackage{cite}
\usepackage{amsmath,amssymb,amsfonts}
\usepackage{algorithmic}
\usepackage{graphicx}
\usepackage{hyperref}
\usepackage{balance}
\usepackage{cleveref}
\usepackage{textcomp}
\usepackage{xcolor}
\usepackage{cleveref}
\usepackage{rotating}
\usepackage{float}
\usepackage{listings}
\usepackage{caption}
\usepackage{tikz}
\usetikzlibrary{automata,positioning}
\usepackage{theorem}
\usepackage{paralist}
\usepackage{multirow}
\usepackage[inline]{enumitem}
\newtheorem{definition}{Definition}

\usepackage{musicography} 
\usepackage{utfsym}

\usepackage{booktabs}        
\usepackage{siunitx}         
\usepackage{array}           
\usepackage[table]{xcolor}   
\definecolor{rowgray}{gray}{0.95}

\definecolor{origbg}{gray}{0.96}
\definecolor{transbg}{gray}{0.92}
\definecolor{brilliantrose}{rgb}{1.0, 0.33, 0.64}
\definecolor{classicrose}{rgb}{0.98, 0.8, 0.91}
\definecolor{gamboge}{rgb}{0.89, 0.61, 0.06}

\newboolean{showcomments}
\setboolean{showcomments}{false} 
\ifthenelse{\boolean{showcomments}}{ 
  \newcommand{\nbnote}[3]{
	\fcolorbox{brilliantrose}{classicrose}{\bfseries\sffamily\scriptsize#1}{
	  \color{#2} 
	  \sffamily
	  \small
	  $\blacktriangleright$
	  \textit{#3}
	  $\blacktriangleleft$
	}
  }
}{
  \newcommand{\nbnote}[3]{}
  
}

\newcommand{\code}[1]{%
  {\fontfamily{lmtt}\fontseries{m}\selectfont
    \codeprocess#1\relax
  }%
}

\newcommand{\new}[1]{{\color{blue}#1}}

\def\codeprocess#1{%
  \ifx#1\relax
  \else
    \ifx#1\_%
      \scalebox{0.5}[1]{\textunderscore}
    \else
      #1%
    \fi
    \expandafter\codeprocess
  \fi
}

\def\BibTeX{{\rm B\kern-.05em{\sc i\kern-.025em b}\kern-.08em
    T\kern-.1667em\lower.7ex\hbox{E}\kern-.125emX}}
\begin{document}

\title{In Perfect Harmony: Orchestrating Causality in Actor-Based Systems\\
\thanks{}
}


\author{\IEEEauthorblockN{Vladyslav Mikytiv}
\IEEEauthorblockA{
\textit{NOVA University Lisbon}\\
Lisbon, Portugal \\
v.mikytiv@campus.fct.unl.pt}
\and
\IEEEauthorblockN{Bernardo Toninho}
\IEEEauthorblockA{
\textit{Instituto Superior Técnico}\\
Lisbon, Portugal \\
bernardo.toninho@tecnico.ulisboa.pt}
\and
\IEEEauthorblockN{Carla Ferreira}
\IEEEauthorblockA{
\textit{NOVA University Lisbon}\\
Lisbon, Portugal \\
carla.ferreira@fct.unl.pt}
}

\maketitle

\begin{abstract}

Runtime verification has gained popularity as a lightweight approach for increasing assurance over systems under scrutiny. 
By performing checks at runtime, it enables dynamic monitoring and alerts for unexpected behavior, thereby improving reliability and correctness.

Actor-based systems present significant challenges for runtime verification. 
Properties frequently span multiple actors with complex causal dependencies, while nondeterministic message interleavings can obscure execution semantics. 
Moreover, most existing monitoring tools are designed for single-process behavior. 
This paper presents ACTORCHESTRA, a runtime verification framework for Erlang that automatically tracks causality across multi-actor interactions.

The framework instruments Erlang systems that comply with OTP guidelines via targeted code injection. 
This method establishes the orchestration infrastructure required to track causal relationships between actors without requiring manual modifications to the target system. 
To ease the specification of multi-actor properties, the framework provides WALTZ, a specification language that automatically compiles properties into executable Erlang monitors that integrate with the instrumented system.
Three case studies demonstrate ACTORCHESTRA's effectiveness in detecting complex behavioral violations in real-world actor systems.
Performance evaluation quantifies the runtime overhead of the monitoring infrastructure and analyzes the trade-offs between added safety guarantees and execution costs.

%
\end{abstract}


\section{Introduction}

As software becomes increasingly entangled with today's society, ensuring
correct system behavior is not only a technical requirement but also an ethical
necessity. Formal methods have emerged to address the need for thorough
verification of software behavior and correctness~\cite{clarke1996formal}.

Reactive systems comprise multiple components that interact via message
passing, posing unique verification challenges~\cite{reactiveComponents}. While
model checking~\cite{clarke1999state} and theorem
proving~\cite{bertot2013interactive} can verify system compliance, they face
significant limitations: model checking suffers from state space
explosion~\cite{leucker2009brief} and theorem proving becomes infeasible due to
system complexity. Testing~\cite{myers2011art}, a widely used validation approach, may miss critical interleavings and testing
scenarios, especially in reactive systems.~\cite{10.1145/1353673.1353681}.

Runtime
verification~\cite{bartocci2018introduction,bauer2011runtime,pnueli2006psl,leucker2009brief}
offers an alternative by monitoring systems during execution. By reasoning over the current execution trace, runtime verification enables compliance checking during execution without constructing the whole state space. A monitor generated from high-level specification languages that define properties of interest runs alongside the system, consuming the trace and issuing verdicts when execution deviates from the specified properties.

Actor-based systems, core to concurrency-oriented
languages~\cite{juric2024elixir} such as
\code{Erlang}~\cite{virding1996concurrent} and \code{Elixir}~\cite{elixir},
exemplify reactive systems in which autonomous entities (actors) interact via
message passing. While the message-centric nature of actors provides a natural
trace definition, monitoring these systems presents fundamental challenges:
\begin{itemize}
  \item{\textbf{Causality correlation}}: How can we track causally related messages 
  across multiple interleaved actor interactions?
  \item{\textbf{Property specification}}: How do we precisely specify properties spanning
  multiple interactions while abstracting away nondeterministic message interleavings?
  \item{\textbf{Transparent Instrumentation}}: Can we inject monitoring capabilities 
  without disrupting system operation or requiring manual code modifications?
\end{itemize}

This work addresses these challenges through a runtime verification framework,
named \code{ACTORCHESTRA}, designed explicitly for \code{Erlang}'s Open Telecom
Platform client-server architectures~\cite{erlangBook, erlangBooktwo}. We focus
on systems that implement the \code{gen\_server} behavior, which captures
common distributed system patterns, including HTTP APIs, service-to-service
communication, and microservices architectures. 

\subsection{Research Questions}
Specifically, this work addresses three research questions: 
\begin{description}[leftmargin=2.5em, style=nextline]
  \item[\textbf{RQ1}] Can causality across multi-actor interactions in OTP-compliant \code{Erlang} systems be tracked automatically? 

  \item[\textbf{RQ2}] Can a specification language support properties spanning multiple actor interactions while abstracting message interleavings? 

  \item[\textbf{RQ3}] What performance overhead is introduced by automated
  causality tracking and runtime monitoring?

\end{description}

\subsection{Contributions}
This work presents \code{ACTORCHESTRA}, a runtime verification framework for actor-based systems following the \code{Erlang} OTP client-server architecture. We make the following contributions:
(1) A \textbf{causality} manager that reconstructs live execution traces and assigns causality tokens to messages across multi-actor interactions;
 (2) A \textbf{compile-time code injector} for \code{Erlang} OTP systems that transparently instruments communication with the causality manager;
 (3) \textbf{WALTZ}, a specification language for expressing causality-aware properties over multi-actor interactions;
 (4) An \textbf{empirical evaluation} of the overhead of our framework
 over three case studies.

\textbf{RQ1} is addressed through the conductor and context injection mechanisms
(\Cref{sub_sec:conductor}, \Cref{sub_sec:contextinjection}), \textbf{RQ2} through
the WALTZ specification language (\Cref{sec_waltz}), and \textbf{RQ3} through
empirical evaluation (\Cref{evaluation}).

\begin{figure*}[t]
  \centering
  \includegraphics[width=0.75\textwidth]{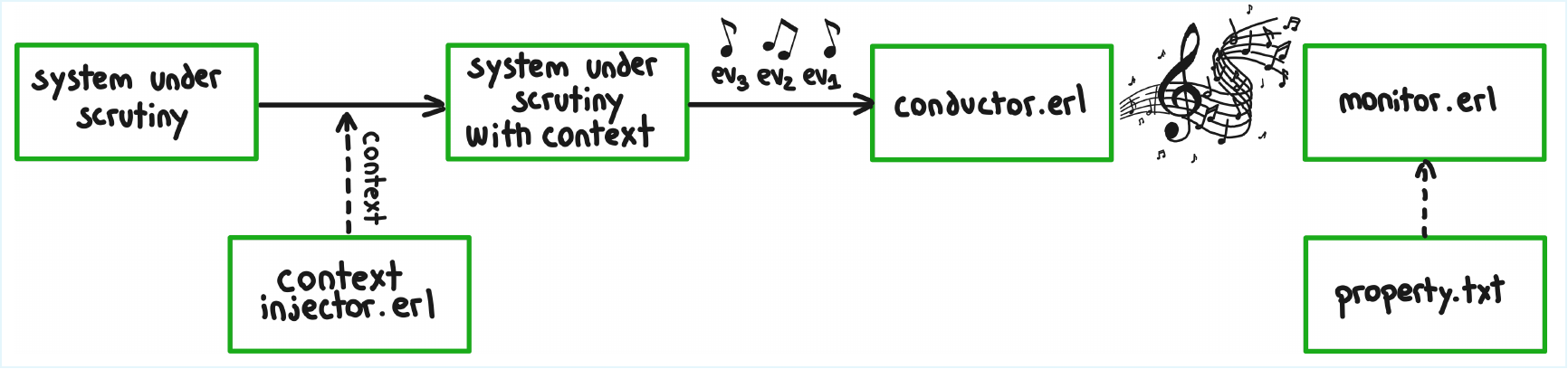}
  \caption{General structure of \code{ACTORCHESTRA}. The \emph{context injector}
  transforms \code{Erlang} source at compile time; the \emph{conductor} intercepts
  all runtime messages, assigns causality tokens, and forwards events to the
  \emph{monitors}, which evaluate WALTZ properties over the resulting causally
  coherent traces.}
  \label{fig:instrumenation3}
\end{figure*}

\section{Background}
\label{background}

This section outlines the foundational concepts supporting our work, covering runtime verification, the actor-based model, causality in concurrent systems, and \code{Erlang}-specific constructs necessary for understanding \code{ACTORCHESTRA}.

\noindent \textbf{Runtime Verification.}
We adopt Leucker's definition~\cite{leucker2009brief} of a \emph{run} as a
possibly infinite sequence of events, where each event is an observable
action from the system's alphabet~\cite{bartocci2018introduction}. A
\emph{trace} is a sequence of events produced during execution. 
In runtime verification, monitors operate over finite prefixes of runs, corresponding to the events observed so far.

A monitor can be automatically synthesized from a high-level formal
specification, interpreting system execution traces to produce
verdicts: \emph{satisfied}, \emph{violated}, or \emph{inconclusive}~\cite{leucker2009brief}.

\noindent \textbf{Actor-based Model.}
The actor model~\cite{agha2001actors} structures computation as independent units (actors) that encapsulate state and behavior and communicate exclusively via asynchronous message passing. 
Languages such as \code{Erlang} and
\code{Elixir}\cite{elixir, erlangBook} implement this model, enabling dynamic actor creation and concurrent execution. Each actor has a unique process identifier (PID), allowing precise tracking of message sender, recipient, and content.

In the context of actor-based systems, the event alphabet corresponds to observable communication events.
The messages exchanged between actors naturally define the trace structure that monitors observe and reason about.
 
\noindent \textbf{Open Telecom Platform.}
The Open Telecom Platform (OTP) provides libraries and design principles for
building concurrent, standardized, and fault-tolerant \code{Erlang}
systems~\cite{erlangBook, erlangBooktwo, hebert2013learn}.
It introduces \emph{behaviors}---generic process patterns for 
common concurrency idioms---where developers implement callbacks while OTP handles supervision, process management, state serialization, and message dispatching.

We focus on the \code{gen\_server} behavior, which encodes client-server
interactions where processes (clients) make calls to a server process and await
responses. Of particular importance is the \code{call} operation: when a client
performs \code{gen\_server:call(ServerPid, Request)}, OTP automatically
sends the request to the server, blocks waiting for the response, and
returns it to the caller when received. The server implements a
\code{handle\_call} callback to process requests, maintaining state
encapsulated within a single process instance, to ensure consistency across all
callbacks. The callback returns a tuple \code{\{reply, Response, NewState\}}
that instructs OTP to send the response back to the waiting client and update
the server's internal state.
Crucially, OTP's automatic reference generation for call operations provides a
foundation for causality tracking. Each \code{gen\_server:call} generates a
unique reference that binds the \code{(request, response)} pair, effectively
implementing a local causality relation. 


\section{Related Work}
\label{related_work}
\label{sec_related_work}

Runtime verification frameworks~\cite{falcone2021taxonomy} share a common objective: examining execution traces to determine compliance with user-defined properties. 
However, most frameworks do not operate over live traces from running systems, and those that do face limitations in expressiveness and automation.

\noindent\textbf{Log-based verification.} MonPoly~\cite{basin2017monpoly} and WHYMON~\cite{limawhymon} provide runtime verification with explainable verdicts. 
Both tools operate over execution traces recorded in log files rather than monitoring live system executions. While both process traces incrementally to approximate online behavior, analysis still occurs over logged data, limiting applicability to highly concurrent and reactive systems. 
A key challenge in runtime verification is the lack of a universally adopted notion of traces~\cite{falcone2021taxonomy}, leading many tools to rely on pre-recorded logs. In contrast, the actor model provides a natural and standard trace through message exchanges, enabling live monitoring.

\noindent\textbf{Session-based monitoring.} Building on multiparty session
types~\cite{multipartysessions}, Fowler proposed
\code{monitored-session-erlang} for \code{Erlang} OTP
applications~\cite{frowler}. The framework introduces causality via session
identifiers that link events within a session, requiring developers to manage
session initiation and switching manually. Participating actors are required to
implement specific directives when using the tool, thereby reducing automation.
While effective for enforcing protocol conformance, the approach is limited as
it cannot reason about message payloads or relationships across interaction
chains.

\noindent\textbf{Process-centric monitoring.}
\code{detectEr}\cite{attard2017runtime} is a runtime verification tool for
\code{Erlang} that leverages native tracing mechanisms and monitors properties
expressed in fragments of $\mu$HML, including recent extensions with data
\cite{adventuresMon}. While \code{detectEr} operates on real execution traces,
it primarily focuses on monitoring individual processes, making it challenging
to express properties spanning multiple actors, such as system-wide invariants.
Although the underlying logic can, in principle, express multi-actor
properties, doing so requires explicit enumeration of all message
interleavings, which becomes intractable in highly concurrent systems. In
practice, the implemented fragment restricts monitoring to a single function
within a single process, preventing direct correlation of causally related
events across larger message chains.

\code{ACTORCHESTRA} addresses these limitations by providing:
\begin{inparaenum}[(1)]
\item automatic causality tracking across multi-actor interactions without developer intervention; 
\item a specification language enabling properties over causal message chains spanning multiple actors;
\item automated instrumentation for OTP-compliant systems.
\end{inparaenum}

\section{ACTORCHESTRA}
\label{actorchestra}

This work proposes \code{ACTORCHESTRA}, a new framework for
runtime verification of actor-based systems following the OTP architecture,
enabling automatic verification of
context-aware properties across multiple components.

\subsection{General Structure}
\label{sub_sec:generalstruct}
Building a runtime verification framework requires addressing two key concerns: 
monitor compilation from specifications and system instrumentation. 
\code{ACTORCHESTRA} comprises three main
entities, depicted in~\Cref{fig:instrumenation3}. Monitors that are deployed without
interfering with system execution and consume enriched traces;  
a conductor that interprets system execution and generates causally coherent traces for monitors; 
and a context injector that modifies the Abstract Syntax
Tree (AST) of \code{Erlang} files at compile-time, adding communication hooks to
the conductor.

Even within actor-based systems, there are multiple different architectures
that a system can follow. Depending on the type of system, causality tracking
between messages changes drastically. In this work, we focus on one
architectural model: client-server interaction. Tackling all the possible
architectures would be nearly impossible. Therefore, we focus on client-server
architectures that follow OTP guidelines and implement the \code{gen\_server}
behavior. In these systems, clients act as entry points, initiating actions and
awaiting responses, while the system handles requests asynchronously from
multiple clients in parallel. Although this imposes client-side sequentiality,
it captures significant distributed system patterns: HTTP APIs,
service-to-service communication, pipelines, and microservices. The defining
characteristic is that every request yields a corresponding response, ensuring
traceability.

\subsection{Running Example}
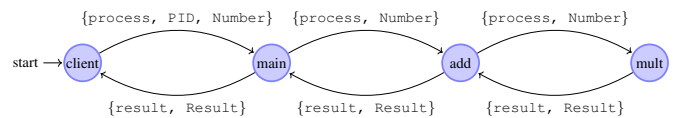
\begin{figure}[b]
\centering
\usetikzlibrary{arrows.meta, positioning, automata, shapes.geometric}

\tikzstyle{nodeB}=[circle,draw=gray!75,fill=blue!20,
                   thick,minimum size=12mm,inner sep=0pt]

\resizebox{\columnwidth}{!}{%
\begin{tikzpicture}[shorten >=1pt, node distance=4.5cm, on grid,
    every state/.style={draw=blue!50,very thick,fill=blue!20}]
    
  \node[nodeB, state, initial] (q1) {client};
  \node[nodeB, state] (q2) [right=of q1] {main};
  \node[nodeB, state] (q3) [right=of q2] {add};
  \node[nodeB, state] (q4) [right=of q3] {mult};

  \path[->] (q1) edge [bend left]  node[above] {\{\texttt{process, PID, Number}\}} (q2);
  \path[->] (q2) edge [bend left]  node[below] {\{\texttt{result, Result}\}} (q1)
                 edge [bend left]  node[above] {\{\texttt{process, Number}\}} (q3);
  \path[->] (q3) edge [bend left]  node[below] {\{\texttt{result, Result}\}} (q2)
                 edge [bend left]  node[above] {\{\texttt{process, Number}\}} (q4);
  \path[->] (q4) edge [bend left]  node[below] {\{\texttt{result, Result}\}} (q3);
\end{tikzpicture}
}
\caption{Simple arithmetic system.}
\label{fig:procadd}
\end{figure}

Consider a simple decentralized arithmetic system in which a \texttt{client}
interacts with a \texttt{main} entry point. Clients send \texttt{connect}
requests and subsequently submit numbers for processing. 
For each input, the system adds 10 and then multiplies the result by 2.
The structure and message signatures are
shown in \Cref{fig:procadd}. A key property to verify is whether the
\texttt{add} process correctly adds 10, i.e., the output value equals the input
plus 10. Additional properties can be defined depending on the desired level of
precision. If the properties are too generic, some specific behaviors may go
undetected. The flexibility of these properties allows the developer to tailor
the verification process to the system's needs. Ultimately, it is the
developer's responsibility to identify and specify the critical properties that
should be monitored at runtime.

\subsection{The $\gamma$ Function}
\label{sub_sec:gammafunction}
We require a function $\gamma$ which is responsible for assigning a context (a
causality token) to a given message. Causally correlated messages will share
the same causality token. One approach is manual management, where developers
control the program's logic and message flow. Crucially, this not only helps
the monitor correlate messages but also ensures proper system flow.  Developers
can use the built-in \code{make\_ref} function to generate unique references
and attach them to message payloads, sharing them across all interactions.  For
trace $\sigma$ with messages \code{m$_1$}, \code{m$_2$}, and \code{m$_3$} from
our running example, it unfolds as follows:
\begin{align*}
  \small\text{\code{m$_1$}} &= \small\text{\code{\{process, <1.0.0>, 10, \#\code{Ref}<1.2>\}}} \\
  \small\text{\code{m$_2$}} &= \small\text{\code{\{process, 20, \#\code{Ref}<1.2>\}}} \\
  \small\text{\code{m$_3$}} &= \small\text{\code{\{result, 40, \#\code{Ref}<1.2>\}}}
\end{align*}
Here, \code{m$_1$} is the \texttt{client}$\to$\texttt{main} request (input~10);
\code{m$_2$} is the \texttt{main}$\to$\texttt{add} request (value~20 after
addition); and \code{m$_3$} is the final result~40 returned after multiplication.
$\gamma(\text{\code{m$_1$}}) \! = \! \gamma(\text{\code{m$_2$}}) \! = \!
\gamma(\text{\code{m$_3$}})= \text{\#\code{Ref<1.2>}}$, since all messages in the chain share the same correlation token. While this achieves precise causality tracking, it couples business logic to token management, increases error likelihood, and weakens OTP's reliability guarantees through ad hoc implementation.

OTP partially addresses this through \code{gen\_server:call}, which automatically
generates a unique reference binding each request to its response:
$\gamma(\small\text{\code{request}}_i, \small\text{\code{response}}_i) =
\small\text{\code{REF}}_i$. {However, this remains pairwise. 
Intuitively, two messages are causally related if one is a direct consequence of the other within the same logical request: in our example, the original client request, the two intermediate forwarding calls, and the final response all belong to the same logical operation. Yet OTP links \code{client}$\to$\code{main} via \code{REF$_1$} and \code{main}$\to$\code{add} via \code{REF$_2$} independently, with no mechanism to record that both belong to the same end-to-end chain. 
A monitor wishing to relate the client's original input to the final computed output across all actors, therefore, cannot do so using OTP references alone. 
A context in \code{ACTORCHESTRA} is a shared token assigned by the conductor and attached to all messages in the same causal chain, making these relationships explicit and available to monitors.} Most importantly, OTP provides a standard infrastructure to build upon, enabling the automation of contexts.

\subsection{The Conductor}
\label{sub_sec:conductor}
The conductor provides automatic end-to-end context management, operating
externally to capture system messages and communicate with monitors. It
serves as an oracle for every system action, ensuring all interactions pass
through it.

Introducing the conductor adds overhead because it becomes 
a coordination point for system communication~\cite{reactiveComponents}. 
However, it is designed for concurrency: it continues accepting incoming messages 
while dispatching outgoing ones with minimal delay. 
Beyond forwarding, the conductor implements the function $\gamma$ by capturing runtime messages 
and maintaining the contexts that relate them. 
This additional processing increases computational cost, but it enables precise causality
tracking and consistent context propagation across all system interactions.

Consider a simple system with two modules: \code{client.erl} and \code{server.erl}.
Clients send messages to the server, which echoes them back. In our
instrumented system, messages from \code{client} to \code{server} are intercepted and
routed through the conductor for monitoring and context management:

\begin{center}
\scalebox{0.95}{
\renewcommand{\arraystretch}{0.95}
\begin{tabular}{c@{\hspace{0.6em}}c@{\hspace{0.6em}}c}
\colorbox{origbg}{\small\code{client\_i} $\rightarrow$ \code{server}}
&
$\Rightarrow$
&
\colorbox{transbg}{%
\begin{tabular}{@{}c@{}}
\small\code{client\_i}\\[-1pt]
$\downarrow$\\[-2pt]
  \small{\code{conductor}}\\[-1pt]
$\downarrow$\\[-2pt]
\small\code{server}
\end{tabular}%
}
\end{tabular}
}
\end{center}

The conductor forwards each message with its associated context to both the
original recipient and the monitor, deciding whether to propagate the existing
context or generate a new one. It implements $\gamma$ as:
\[
\small
\gamma(m)=
\begin{cases}
  \texttt{Context} & \text{if } \texttt{m}.\texttt{context} \neq \texttt{undefined} \\
  \texttt{gen\_context()} & \text{otherwise.}
\end{cases}
\]

The conductor handles this via a \code{call} endpoint that all processes
invoke, sharing their context. The handler inspects messages for existing
contexts and reuses or generates as needed (Listing~\ref{contextconductor}).
\begin{lstlisting}[caption={Conductor context assignment.},label={contextconductor},language=Erlang,keywordstyle=\color{blue},basicstyle=\ttfamily\footnotesize]
call(To, Msg, undefined, Mod) ->
  Context = make_ref(),
  gen_server:call(conductor, 
  {send, To, Msg, Context, Mod});
call(To, Msg, Context, Mod) ->
  gen_server:call(conductor, 
  {send, To, Msg, Context, Mod}).
\end{lstlisting}

Next, the conductor's internal handler is invoked, responsible for reporting
the occurrence of this event to the monitor (Listing
\ref{sendcontextconductor}), capturing the response, and redirecting the
message to the original recipient. Asynchrony is essential to prevent
deadlocks. Since all processes communicate through the conductor, synchronous
processing could cause the conductor to block when a response depends on
another process's action. By spawning worker processes (Listing
\ref{sendcontextconductor}) for message redirection, the conductor avoids
deadlocks, especially critical under high concurrency with multiple clients.
The spawned worker redirects the message, captures the response when it comes,
and reports back to the monitor.
\begin{lstlisting}[caption={Conductor message redirection.},label={sendcontextconductor},language=Erlang,keywordstyle=\color{blue},basicstyle=\ttfamily\footnotesize]
handle_call({send, To, Msg, Context, Mod},
            From, State) ->
  ToMod = get_target_module(To),
  monitor ! {Mod, ToMod, Msg, Context},
  spawn(fun() ->
    worker(To, Msg, Context, Mod, From, ToMod)
  end),
  {noreply, State}.

worker(To, Msg, Context, Mod, From, ToMod) ->
  Reply = gen_server:call(To, 
    {with_context, Context, Msg}),
  FinalReply = case is_client_module(Mod) of
    true -> {Reply, Context};
    false -> Reply
  end,
  monitor ! {ToMod, Mod, Reply, Context},
  gen_server:reply(From, FinalReply).
\end{lstlisting}

Since clients serve as communication boundaries, they do not know the generated
context value until receiving a response. Therefore, the context must be
propagated back to clients so they can store it locally for subsequent
requests. To accomplish this, the conductor checks whether the response is
being sent to a client module (Listing \ref{sendcontextconductor}). When a
client module is detected, the conductor modifies the response message to
include the context value, enabling the client to cache it for future requests.
This client-based propagation supports the system's architectural strategy of
establishing contexts at client boundaries. Treating each client interaction as
a causal chain's starting point provides a natural entry point for trace
interception and enables session awareness. While all requests from a client
share the same context internally, the monitor interprets message semantics to
enforce logical isolation between requests, allowing it to identify separate
causal chains within shared contexts.

\subsection{Context Injection Mechanism}
\label{sub_sec:contextinjection}
To route the communications through the conductor and accept special
messages from it, we implement an automated context injector that modifies the
system's components AST at compile time, adding communication hooks. Leveraging
OTP allows for minimal changes with strong safety assumptions. The
injector performs four primary transformations: state management enhancement
with a context field, context reception handling, message call transformation,
and context extraction.

To enable these injections, \code{Erlang} modules must include
the header:
  {\small\text{\code{-compile(parse\_transform, context\_injector)}}}.

This flag instructs the \code{Erlang} compiler to process the module's AST according
to the rules defined in \code{context\_injector.erl}, which analyzes
the existing code structure and applies the required modifications
automatically.

In the OTP framework, a process's state is formally an
opaque value passed through successive callback invocations. OTP does not
restrict its type: any \code{Erlang} term is admissible. In principle, a
process may maintain its state as a single integer, a tuple, or even a list.
However, both the \code{Erlang}/OTP standard library and idiomatic
application design show that structured representations, specifically records
and maps, are overwhelmingly preferred in practice~\cite{erlangBook,
erlangBooktwo}.

The injector supports both record-based and map-based state management,
following OTP conventions. For record-based modules, it augments the record
definition with a \code{context} field initialized to \code{undefined}. For
map-based modules, context operations use \code{maps:put} and
\code{maps:get}.

The key modification transforms every \code{call} request. Instead of sending
calls directly to the original target, they are redirected to the conductor
with the original information and metadata. By intercepting all messages, the
conductor assigns context references and maintains them across causal chains.
To illustrate, consider a standard OTP-based request:
\begin{align*}
  \small\text{\code{gen\_server:call(Target, Msg)}}
\end{align*}

Through parse-time code transformation, via the \code{parse-transform} module
that \code{Erlang} offers, our context injector automatically enhances the
code to redirect the original call to the conductor, sharing the context:
\begin{align*}
  \small\text{\code{conductor:call(Target, Msg, Context, ?MODULE)}}
\end{align*}

These transformations establish the hooks with the conductor and
share an internal context value. 

Each \code{gen\_server} is augmented with a dedicated \code{handle\_call}
clause to receive and store context updates from the conductor. To
ensure atomicity, these updates are processed as a single message. 
The \code{Context} value is updated for each interaction, being rewritten
constantly through its lifetime. As long as the original message and its
associated context are communicated to the monitor as a unified pair by the
conductor, causal consistency is maintained. This process is
implemented through the \code{with\_context} handler, which receives both the
original request and the context from the conductor, updates the state
accordingly, and executes the original function within the same module. An
example implementation of this handler is shown in Listing~\ref{withcontext}.
\begin{lstlisting}[caption={Component context extraction.},label={withcontext},language=Erlang,keywordstyle=\ttfamily, basicstyle=\ttfamily\footnotesize]
handle_call({with_context, Ctxt, Msg}, From, State):
  % Map-based:
  StateWithCtxt = maps:put(context, Ctxt, State),
  % Record-based:
  StateWithCtxt = State#state{context = Ctxt},
  ?MODULE:handle_call(Msg, From, StateWithCtxt);
\end{lstlisting}

This architecture ensures context flows throughout the interaction chain. 
The conductor forwards all events to the monitor while distributing context 
to components through the injected handler. As the context propagates, 
each \code{with\_context} handler updates its process's context, 
maintaining causal consistency across the entire request-response chain.

Client processes must update their context field. Since clients cannot
automatically update during the call, context is piggybacked on the response.
Client modules extract context from results and update state accordingly, 
enabling correlation of subsequent requests from the same client when needed. 
This augmentation is performed only on the client side.

Special care must be taken when propagating context. Each actor has a
\code{context} field that can be overwritten. To maintain causal consistency,
each message must be associated with the correct context before being shared
with the conductor. Due to asynchrony, the context should be extracted prior to
sending the message to the conductor to prevent spawned processes from
accessing incorrect or outdated variables. To prevent race conditions in which
concurrent requests overwrite the context field, the injector adds a single
line at the beginning of each \code{handle\_call} function to immediately
capture the current context value into an immutable variable.  This ensures
that any spawned processes use the correct context for their messages,  even if
subsequent requests update the state.

The proposed solution introduces a wrapper-based architecture that
changes how context propagation occurs in inter-process
communication. The conductor embeds contextual information directly within the
message payload using a specialized envelope pattern, sending it to a special
endpoint injected into each file.
This allows context to propagate 
through the chain of interactions, then be captured by the conductor and shared
with the monitor, making it possible to establish causal relationships between
message interaction inside the system that are part of the same logical
requests.

\section{Waltz}
\label{sec_waltz}

Existing specification languages for actor-based monitoring either cannot
capture properties spanning multiple actor interactions or lack causality
awareness. We require a specification language integrated with the actor model
that enables context-aware multi-actor properties while reducing specification
complexity. WALTZ addresses these needs as a domain-specific language that
compiles to executable \code{Erlang} monitors. Research has shown that
specification languages closer to the domain reduce cognitive load and error
rates~\cite{domainlanguage}. Additionally, a domain-specific language can offer
more effective error messages, enhanced debugging support, and seamless
integration with actor-based development tools. 
This philosophy guides the design of WALTZ. 
As analyzed in~\cite{falcone2021taxonomy}, developing dedicated specification 
languages is standard practice in runtime verification, 
with most RV tools following this approach.

\subsection{The WALTZ Language}
WALTZ is built around two core concepts: interaction signatures (to match
message exchanges between actors) and data constraints (boolean constraints
over payload values), resulting in a code-like specification style.
A WALTZ specification is
evaluated over contexts. A context represents a causally coherent interaction
chain reconstructed by the conductor (\Cref{actorchestra}). Then we can
verify properties over such chains, each in its own context, across multiple
actor interactions, without worrying about all possible interleavings.

\begin{definition}
The grammar of WALTZ formulas is:
{\small
\[
\renewcommand{\arraystretch}{1.12}
\setlength{\arraycolsep}{8pt}
\begin{array}{@{}l@{\;::=\;}l@{\qquad}l@{\;::=\;}l@{}}
\multicolumn{4}{@{}l@{}}{\varphi ::= \Omega(\varphi)\mid\Theta(\varphi)\mid\varphi;\varphi\mid\alpha:\delta}\\[3pt]
\delta & \top\mid\bot\mid c\mid\lnot\delta
& \alpha & \texttt{send}_{\texttt{id}\rightarrow\texttt{id}}\!\left\{\texttt{M}\right\}\\[3pt]
\texttt{M} & \texttt{T}\mid\texttt{T},\texttt{M}\mid\left\{\texttt{M}\right\}
& \texttt{T} & \texttt{cons}\mid\texttt{var}
\end{array}
\]
}
\end{definition}

The core construct is $\alpha : \delta$, where $\alpha$ is an interaction signature (a message pattern between two actors), and $\delta$ is a boolean constraint, indicating that messages $\alpha$ must satisfy $\delta$. 
To capture unconstrained messages, we write $\alpha : \top$.

A signature specifies the operation type, source and destination actors, and
message structure. For example, \texttt{$\text{send}_{A \rightarrow B}$ \{X, Y,
\{Z, ok\}\}} matches messages from actor A to actor B with the specified
payload structure. Uppercase identifiers are variables bound at runtime; lowercase identifiers are fixed atoms matched literally. For now, we only support the \code{send} event, as it suffices to capture a wide range of properties.

The chain operator (\texttt{;}) sequences formulas. The expression $\alpha_1 \ {:} \ 
\delta_1 \ {;} \ \alpha_2 \ {:} \ \delta_2$  specifies that a message matching
$\alpha_1$ and satisfying $\delta_1$ must first be observed, followed by a message
matching $\alpha_2$ and satisfying $\delta_2$. Chains are unbreakable: if any constraint fails, the chain is considered broken, and depending on the modal operator wrapping the property, this may lead to a violation.

WALTZ introduces modal operators for reasoning about contexts. $\Omega$ 
checks whether a property is satisfied in all system contexts,
whereas $\Theta$ checks whether at least one context satisfies the property.
A well-formedness function validates whether formulas are correctly constructed,
specifically, whether a given formula is encapsulated within a modal operator. 
Only well-formed formulas are considered valid.

\subsection{Semantics}

An interaction with the system produces a trace composed of various events,
some of which are causally or logically related. We refer to such related
events as belonging to the same context. A trace may contain multiple contexts,
each representing an independent chain of causally related events. Our goal is
to verify properties over these individual contexts. Since event ordering is
not always guaranteed and interactions may overlap, we assign a context
identifier to each event to determine which events belong together. 

Formally, a message will be mapped to a context that represents a causal chain
in the system. Let $\mathcal{C}$ be the set of all context identifiers (e.g.,
$\Delta_0$, $\Delta_1$), we introduce a function $\gamma : \Sigma \rightarrow
\mathcal{C}$ that maps each event (a message) in the trace to a unique context
identifier. Intuitively, $\gamma(\sigma_i)$ tells us to which context the event
$\sigma_i$ belongs.

$\Delta_0$ denotes the main context, which always exists, and additional
contexts may be introduced as a system evolves. 
Monitors interpret contexts depending on how properties are specified. 
Contexts can be structured hierarchically, allowing modeling of nested behaviors.
We define a hierarchical relation between two contexts, written 
${\Delta_b \blacktriangleright \Delta_a}$, denoting that context
$\Delta_b$ is derived from context $\Delta_a$. This hierarchical relationship
captures the causal or structural dependency between contexts.

\begin{definition}
  Let $\sigma=\sigma_1 \sigma_2 ... \in \Sigma^\omega$ be an
  infinite sequence of events, $\mathcal{C}$ the set that represents
  the contexts and the relations between them in the given trace, $\gamma$ the
  function that maps messages to a context, $\Delta$ the current
  context, and $i \leq j$. The semantics of WALTZ is defined as follows:
\end{definition}
\begin{small}
\begin{align*}
  \sigma, \Delta, i, j \;&\models\; \alpha : \delta \hspace{-1.4em} &  
  	\text{ iff } & \exists_{k \geq i \wedge k \leq j} : \sigma_k \models \delta \wedge \sigma_k \bowtie \alpha \wedge \gamma(\sigma_k) = \Delta  \\
  \sigma, \Delta, i, j \;&\models\; \varphi \; ; \; \varphi' \hspace{-1.4em}  & \text{ iff } & \exists_{k \geq i \wedge k < j} : \sigma, \Delta, i, k \models \varphi \wedge  \sigma, \Delta, k, j \models \varphi' \\
  \sigma, \Delta, i, j \;&\models\; \Omega (\varphi) \hspace{-1.4em}  & \text{ iff } & \forall_{\Delta_{n} \blacktriangleright \Delta} : \sigma, \Delta_{n}, i, j \models \varphi \\
  \sigma, \Delta, i, j \:&\models\; \Theta (\varphi) \hspace{-1.4em}  & \text{ iff } & \exists_{\Delta_{n} \blacktriangleright \Delta} : \sigma, \Delta_{n}, i, j \models \varphi
\end{align*}
\end{small}
where $\bowtie$ is the operator that checks if a message has a given signature,
and $\blacktriangleright$ is the operator that relates a sub-context with a
higher one in the hierarchy. The satisfaction relation $\sigma, \Delta, i, j
\models \varphi$ denotes that formula $\varphi$ holds in trace $\sigma$
under context $\Delta$, between $\sigma_i$ and $\sigma_j$. This reasoning is
necessary to clearly separate message chains.

Intuitively, given a trace $\sigma$ and the existing contexts $\mathcal{C}$,
we can state: $\sigma, \Delta, i, j \models \alpha : \delta$ means
that three conditions must hold simultaneously. First, a message with signature
$\alpha$ must be observed in the trace $\sigma_i \ ... \ \sigma_j$. Second, this message
must satisfy the constraint $\delta$. Finally, the message must belong to the
context $\Delta$ in which it is being evaluated. 
The property is satisfied only when all three conditions hold.

We say that $\sigma, \Delta, i, j \models \varphi ; \varphi'$ if there exists
a sub-trace $\sigma_i \ ... \ \sigma_k$ in which a message satisfies $\varphi$,
and another sub-trace $\sigma_k \ ... \ \sigma_j$  that satisfies $\varphi'$,
where $i \leq k \wedge k < j $. \mbox{The operator \texttt{;} } allows the
definition of message chains by enforcing an order between events and
specifying the boolean constraints that each message in the chain must satisfy.
Importantly, all events in the same chain belong to the same context.

The $\Omega$ and $\Theta$ operators are the most nuanced constructs in the language, as they
introduce the evaluation over different contexts. Formally, a trace
$\sigma$ under the set of contexts $\mathcal{C}$, and current context $\Delta$,
satisfies $\Omega(\varphi)$ if, for all contexts $\Delta_n$ such that $\Delta_n
\blacktriangleright \Delta$ (i.e., $\Delta_n$ was derived from $\Delta$), the
formula $\varphi$ holds in each derived context. In short, $\Omega(\varphi)$
universally quantifies over all sub-contexts forked from the current one,
requiring that $\varphi$ be satisfied in each.
The $\Theta$ operator works analogously to the $\Omega$ but with
existential semantics: instead of verifying the property across all possible contexts, it succeeds if there exists at least one context in which the property holds.

\subsection{Evaluation and Tracing}
\begin{figure}[t]
\centering
\usetikzlibrary{arrows.meta, positioning, automata, shapes.geometric}
\scalebox{0.6}{%
\begin{tikzpicture}[shorten >=1pt,on grid,
    every state/.style={draw=blue!50,very thick,fill=blue!20}]
  \tikzstyle{nodeB}=[circle,thick,draw=gray!75,fill=blue!20,minimum size=14mm]
  
  \node[nodeB, state] (q_000) at (0,0) {$\Delta_0$};
  \node[nodeB, state] (q_222) at (-2,-2) {$\Delta_1$}; 
  \node[nodeB, state] (q_333) at (0,-2) {$\Delta_2$};
  \node[nodeB, state] (q_111) at (2,-2) {$\Delta_3$};
  
  \path[->] (q_000) edge node [above left] {} (q_222);
  \path[->] (q_000) edge node [right] {} (q_333); 
  \path[->] (q_000) edge node [above right] {} (q_111);
\end{tikzpicture}
}
\caption{Context tree example.}
\label{latticecontexttheta}
\end{figure}
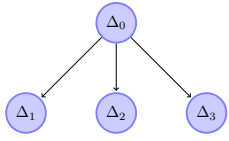

Although a trace may be infinite,
monitors process only messages 
matching user-specified signatures that form the specified chain
within each context, inducing a finite sub-trace. 
This structure forms a tree (our context tree $\mathcal{C}$), with evaluation performed over leaf nodes. 
For clarity, we illustrate only one hierarchy level.
The language supports multiple nesting levels of contexts. 
\Cref{latticecontexttheta} shows an example where all
messages are filtered by a user-specified signature $\alpha$.

Depending on the modal operators used ($\Omega$ or $\Theta$), we traverse all the contexts that exist in the tree. For $\Omega$, we look for a violation in one context; otherwise, we keep monitoring. For $\Theta$, we aim to achieve satisfaction; if none is found, we continue monitoring. Thus, we stop only when a violation occurs (for $\Omega$) or when satisfaction occurs (for $\Theta$).
While these operators support nesting for more complex specifications, we defer a complete exposition of the specification language, focusing here on the monitoring architecture.

For our running example, verifying that for every
client the \code{add} process correctly adds 10 can be defined as
property $\varphi$:
\begin{lstlisting}[
  language=Erlang,
  basicstyle=\ttfamily\footnotesize,
  mathescape=true
]
    $\varphi$ = $\Omega$(send$_{\text{\texttt{main}} \rightarrow \text{\texttt{add}}}$ {process, _, Number1} : $\top$;
            send$_{\text{\texttt{add}}  \rightarrow \text{\texttt{mult}}}$ {process, Number2} :
              Number2 = Number1 + 10)
\end{lstlisting}

This property captures the first message (binding \code{Number1}), 
then the second message (binding \code{Number2}), 
and relates the two variables to check a boolean constraint within the same context.
Encapsulated by $\Omega$, it repeats this verification across all contexts.
WALTZ handles causality tracking and interleaving automatically,  so developers need {to} specify only the logical property.

\subsection{Monitor Compilation}
Monitors for both $\Omega$ and $\Theta$ share a common structural pattern. To
enforce the sequential observation of events within a context, each monitor is
compiled into a chain of nested \code{receive} statements that consume
messages from the mailbox in the prescribed order, as illustrated in
Listing~\ref{receivechain}.
\begin{lstlisting}[caption={Message chain consumption pattern.},label={receivechain},language=Erlang,keywordstyle=\color{blue},basicstyle=\ttfamily\footnotesize]
receive Signature1 -> if BooleanConstraint1 ->
    receive Signature2 -> if BooleanConstraint2 ->
        % Continue chain
\end{lstlisting}

For $\Omega$, the monitor verifies boolean constraints at each step, signaling a violation when one fails. 
For $\Theta$, the monitor tracks whether any context satisfies the property and performs similar checks.
For example, the monitor compiled for property $\varphi$ is depicted in Listing \ref{generated}. 
The monitor begins by consuming one context, anchoring that value at the beginning of the receive chain.
Since we are monitoring for $\Omega$, at each step we evaluate the boolean constraint for violations. When the second message in the chain
arrives, we can relate previously established variables to create our properties, showcasing WALTZ's strength in relating multiple variables across
chains of messages. Depending on the properties we define, they might be more fine-grained or more generic, allowing different levels of fault localization.

\begin{lstlisting}[caption={Monitor for property $\varphi$.},label={generated},language=Erlang,keywordstyle=\color{blue},basicstyle=\ttfamily\footnotesize]
loop() ->
  receive
    {main, add, {process, _, Num1}, Context} ->
      if true ->
        receive
          {add, mult, {process, Num2}, Context} ->
            Constraint = (Num2 == Num1 + 10),
            if Constraint -> loop()
            ; true -> violated
            end
        end
      ; true -> violated
      end
  end.
\end{lstlisting}

\section{Evaluation}
\label{evaluation}

We evaluate \code{ACTORCHESTRA} on three systems: a synthetic arithmetic pipeline, a
complex chat room application, and \code{Lasp}~\cite{lasp}, a production CRDT
implementation. Our goal is to quantify instrumentation overhead and understand
trade-offs between runtime verification and unmonitored execution.

\begin{table*}[t]
  \caption{Performance impact across \code{Lasp} and chat room systems. OH = Overhead (execution time and latency), $\Delta$ Th = Throughput degradation from baseline. Representative configurations shown for each system (light, medium, and heavy loads).}
  \label{table:consolidated_performance}
  \centering
  \footnotesize
  \setlength{\tabcolsep}{3pt}
  \renewcommand{\arraystretch}{1.15}
  \sisetup{
    round-mode=places,
    round-precision=1,
    table-number-alignment=center
  }
  \begin{tabular}{
    l
    r
    S[table-format=5.1]
    S[table-format=5.1]
    S[table-format=3.1]
    S[table-format=4.1]
    S[table-format=4.1]
    S[table-format=3.1]
    S[table-format=4.1]
    S[table-format=4.1]
    S[table-format=2.1]
  }
    \toprule
    \multirow{2}{*}{\textbf{System}} & 
    \multirow{2}{*}{\textbf{Config}} & 
    \multicolumn{3}{c}{\textbf{Execution Time (ms)}} &
    \multicolumn{3}{c}{\textbf{Latency (ms)}} &
    \multicolumn{3}{c}{\textbf{Throughput (m/s)}} \\
    \cmidrule(lr){3-5} \cmidrule(lr){6-8} \cmidrule(lr){9-11}
    & & {Base} & {Instr.} & {OH (\%)} & {Base} & {Instr.} & {OH (\%)} & {Base} & {Instr.} & {$\Delta$ (\%)} \\
    \midrule
    \multirow{6}{*}{Chat Room}
    & 5C$\times$30M   & 38.3    & 75.7    & 97.7  & 1.3    & 2.5    & 98.3  & 4601.4 & 2079.0 & 54.8 \\
    & 10C$\times$30M  & 64.8    & 140.5   & 116.7 & 2.1    & 4.6    & 117.7 & 4698.1 & 2160.1 & 54.0 \\
    & 20C$\times$30M  & 115.6   & 267.1   & 131.1 & 3.8    & 8.8    & 131.4 & 5251.1 & 2263.8 & 56.9 \\
    & 50C$\times$30M  & 267.1   & 635.0   & 137.7 & 8.8    & 20.9   & 137.9 & 5631.3 & 2364.9 & 58.0 \\
    & 100C$\times$30M & 476.8   & 1165.9  & 144.6 & 15.7   & 38.2   & 143.8 & 6307.6 & 2576.2 & 59.2 \\
    & 200C$\times$30M & 896.6   & 2265.3  & 152.7 & 29.4   & 74.1   & 151.8 & 6696.0 & 2649.1 & 60.4 \\
    \midrule
    \multirow{6}{*}{Lasp}
    & 100C$\times$10M  & 640.4   & 1205.4  & 88.2  & 63.9   & 120.4  & 88.2  & 1571.3 & 905.3  & 42.4 \\
    & 200C$\times$10M  & 1267.7  & 2032.2  & 60.3  & 126.5  & 202.6  & 60.3  & 1590.3 & 1014.8 & 36.2 \\
    & 500C$\times$10M  & 3972.2  & 4687.8  & 18.0  & 396.6  & 468.0  & 18.0  & 1313.1 & 1088.7 & 17.1 \\
    & 1000C$\times$10M   & 6778.0  & 8571.0  & 26.4  & 676.7  & 855.7  & 26.4  & 1490.2 & 1181.0 & 20.8 \\
    & 1500C$\times$10M & 11460.3 & 12845.9 & 12.0  & 1144.0 & 1282.2 & 12.0  & 1325.1 & 1180.8 & 10.9 \\
    & 2000C$\times$10M   & 13612.1 & 16374.9 & 20.2  & 1358.7 & 1633.7 & 20.2  & 1473.9 & 1225.2 & 16.9 \\
    \bottomrule
  \end{tabular}
\end{table*}

\subsection{Experimental Setup}
All experiments used deterministic workloads (unaltered)
and instrumented (with \code{ACTORCHESTRA}) versions. We randomize execution
order to reduce bias, with 10 repetitions per configuration. The experiments
were conducted on a MacBook Air with an M1 chip [8GB RAM] and a Windows
Machine with an Intel i7-10700 processor [32GB RAM]. To ensure reliable measurements,
we account for \code{Erlang} VM warmup, as the lazy start
architecture~\cite{warmup} can introduce artifacts. Warmup requests stabilize
performance before actual experiments begin.

We measure three indicators: 
\begin{enumerate*}[label=(\arabic*)]
\item overhead percentage (relative execution time
increase from instrumentation);
\item latency (average request processing time);
and 
\item throughput (requests per second). These metrics assess
\code{ACTORCHESTRA}'s impact across architectural patterns and workloads. The
measurements exclude setup time and measure only active operation time. 
\end{enumerate*}
Due to the
running example's simplicity, we list only the time overhead; full results are
included in our replication package~\cite{replication}.

\subsection{Evaluated Systems}
For each system, we evaluated different properties. In the arithmetic system,
we use the property $\varphi$ listed previously. For the chat system, we verify
that all client messages appear only in the chat rooms they joined. For
\code{Lasp}, we verify that the counter CRDT~\cite{crdts} increments after each
client update request. Table~\ref{table:consolidated_performance} summarizes
performance across representative configurations for the chat room and
\code{Lasp} systems, where 5C $\times$ 30M denotes 5 clients each issuing 30
messages.

\subsection{Simple Arithmetic System}
The arithmetic system implements a pipeline architecture, a classic pattern in \code{Erlang} systems. Although individual clients wait for responses before issuing subsequent requests, the system handles multiple clients asynchronously by spawning dedicated processes for each request. 

We evaluated the system with 1 to 16 concurrent clients, issuing different configurations of requests. Table~\ref{table:multi_client_colored}
summarizes the execution-time overhead for the highest-load configuration per client. The instrumented version exhibits 105--130\% overhead, with the overhead increasing slightly as client concurrency grows, demonstrating that the conductor's asynchronous design scales reasonably with concurrency, though architectural bottlenecks emerge at higher client counts.

\begin{table}[t]
  \caption{Execution time overhead arithmetic system.}
\label{table:multi_client_colored}
  \centering
  \footnotesize
  \setlength{\tabcolsep}{2.8pt}
  \renewcommand{\arraystretch}{1.0}

  \sisetup{
    round-mode=places,
    round-precision=1,
    table-number-alignment=center
  }

  \begin{tabular}{
    r
    r
    r
    S[table-format=3.2]
    S[table-format=4.2]
    S[table-format=3.1]
  }
    \toprule
    {Clients} & {Req./Cl.} & {Total} &
    {Base (ms)} & {Instr. (ms)} & {OH (\%)} \\
    \midrule
     1 & 10K &  10K & 102.36 &  209.27 & 105.4 \\
     2 & 10K &  20K & 202.77 &  420.83 & 108.1 \\
     4 & 10K &  40K & 318.02 &  675.57 & 113.1 \\
     8 & 10K &  80K & 454.58 & 1031.92 & 128.5 \\
    16 & 10K & 160K & 945.26 & 2166.70 & 129.6 \\
    \bottomrule
  \end{tabular}
\end{table}

The arithmetic system demonstrates that our instrumentation imposes moderate overhead, roughly doubling execution time. The overhead increases modestly with load, remains bounded, and remains manageable even at high request volumes, validating that the conductor's asynchronous architecture handles concurrent message streams without becoming pathological.

\subsection{Chat System}
We implemented a chat system with three components: a \code{chat\_server} that handles client communications, \code{chat\_room} actors that manage individual rooms, and a \code{registry\_server} for client registration.
Clients join rooms, post, and disconnect, simulating real chat interactions.

\begin{figure}[t]
  \centering
  \includegraphics[width=\columnwidth]{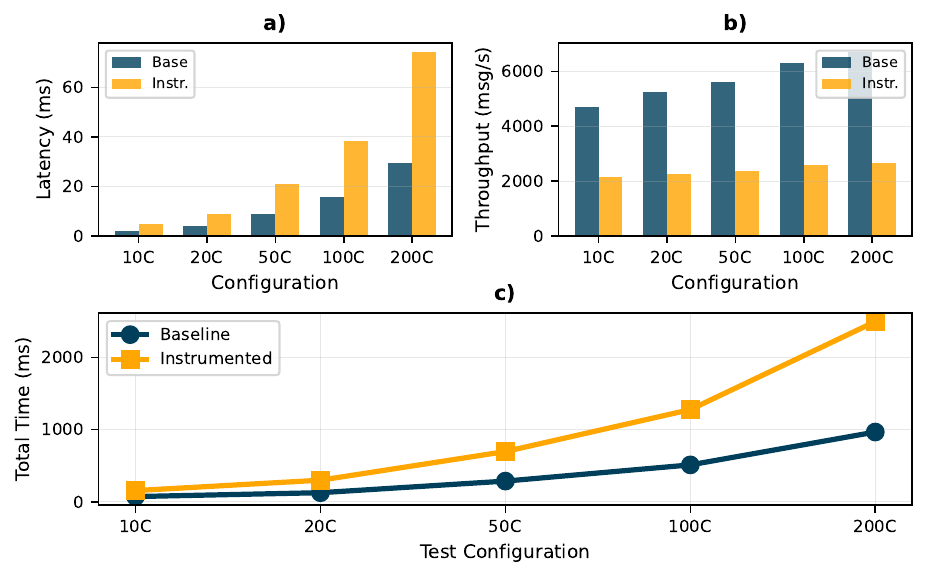}
  \caption{a) Chat System Latency. b) Chat System Throughput. c) Chat System Execution Time.}
  \label{fig:chat_three_combined}
\end{figure}

\paragraph{Execution Time Overhead}
Table~\ref{table:consolidated_performance} and Figure~\ref{fig:chat_three_combined} present message processing overhead across six client configurations. Overhead is bounded between 98--153\%. The configuration with 200 clients shows the highest overhead (153\%), showing
a growing trend as the number of clients increases in the system.

The chat system exhibits higher  overhead than the arithmetic system (98--145\% vs. 105--130\%), largely due to its greater architectural complexity.
Each client action triggers multiple instrumented operations, including connection handshakes, room membership management, message-log persistence, and durable state updates, all coordinated and routed through the conductor, thereby contributing to the increased overhead.

\paragraph{Latency Analysis}
Table~\ref{table:consolidated_performance} and Figure~\ref{fig:chat_three_combined} present per-message latency, measured during the active messaging phase to isolate actual system work from setup overhead. The instrumented version exhibits a latency overhead of 98--152\% (i.e., roughly 2.5$\times$ baseline latency) across all configurations.

\paragraph{Throughput Analysis}
Table~\ref{table:consolidated_performance} and Figure~\ref{fig:chat_three_combined} characterize system capacity under instrumentation. The baseline achieves 4,600--6,600 messages per second (msg/s), while the instrumented version sustains 2,000--2,700 msg/s, a degradation of 55--60\%. Despite this drop, the instrumented system's capacity of 2,000+ msg/s
remains substantial for development-time evaluation, supporting realistic scenarios such as concurrent multi-user sessions with dozens of active clients, stress testing, and integration testing of complex message flows. The degradation is stable across scales, enabling predictable test planning, though it raises the concern that throughput could drop further as the number of clients and requests increases.


The chat system evaluation demonstrates that our approach effectively handles realistic distributed complexity. Execution time overhead stabilizes at
128--145\% across 40$\times$ client scaling, latency overhead ranges from 98\% to 150\% as concurrency increases (before rising at extreme scale), and throughput degradation remains bounded at 55--60\%. The higher overhead compared to the arithmetic system (128--145\% vs. 105--130\%) reflects the architectural complexity of real distributed systems, having more message coordination performed by the conductor.

\subsection{Lasp: Industrial Case Study}
To validate our approach on a production-grade system, we evaluated \code{Lasp}~\cite{lasp}, an industrial implementation of Conflict-Free Replicated Data Types (CRDTs)~\cite{crdts} for \code{Erlang} for eventually consistent distributed systems. \code{Lasp} provides multiple CRDT types and supports multi-node deployments, in which nodes can join a service, consult the shared state, and perform updates that propagate asynchronously across replicas. We instrumented \code{Lasp} by creating a \code{client.erl} endpoint that communicates with the \code{Lasp} backend to perform operations on a \code{state\_gcounter}, a grow-only counter CRDT. Our benchmark scenarios involve multiple concurrent clients incrementing this counter, with monitors verifying that observed values increase monotonically after each update, testing both correctness and performance under instrumentation.

\begin{figure}[t]
  \centering
  \includegraphics[width=0.9\linewidth]{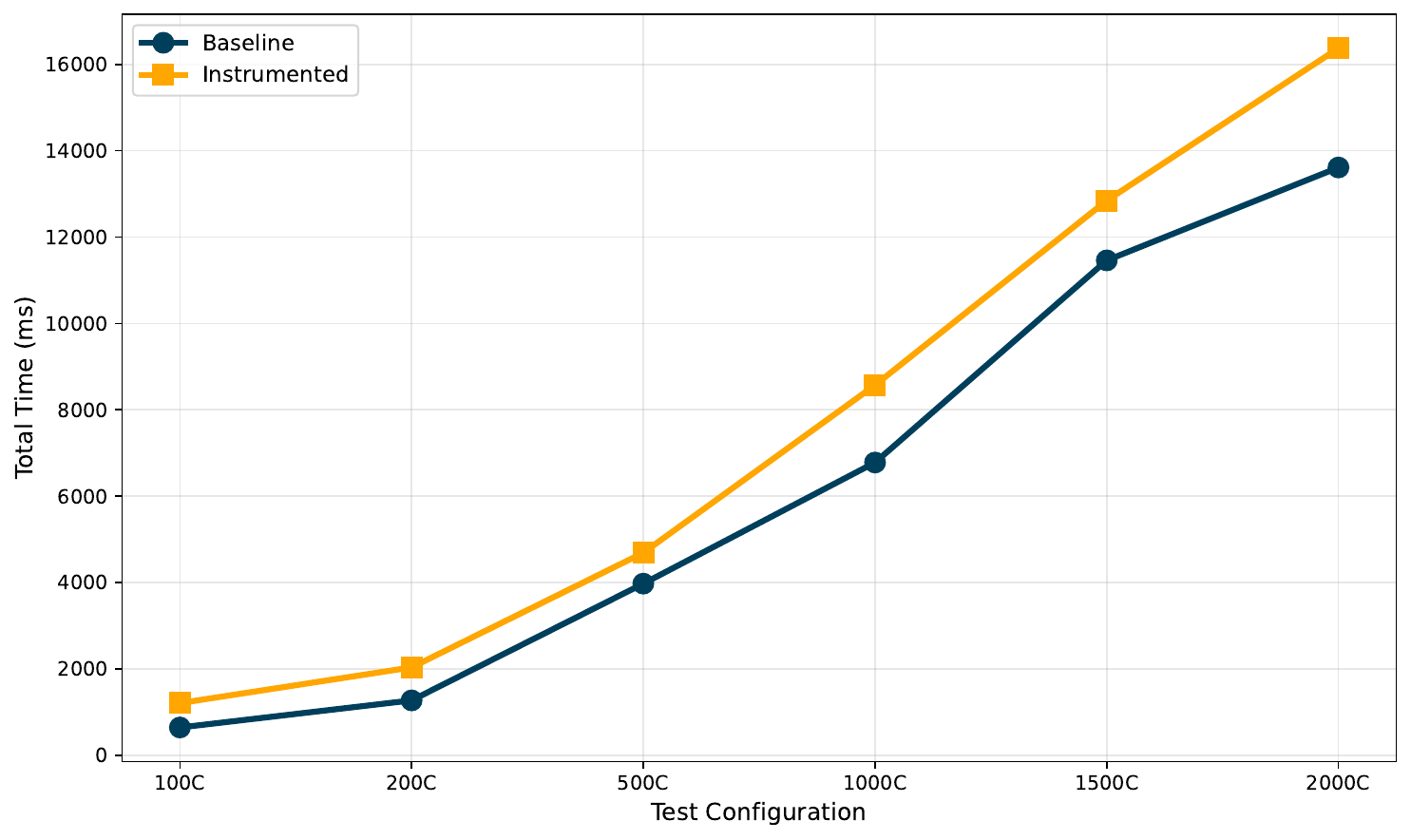}
  \caption{Execution time comparison for Lasp.}
  \label{fig:lasp_time}
\end{figure}

\paragraph{Execution Time Overhead}
Table~\ref{table:consolidated_performance} and Figure~\ref{fig:lasp_time} report execution-time overhead across six client configurations. 
Notably, the overhead \emph{decreases} as the system scales: from 88.2\% at 100 clients to a minimum of 12.0\% at 1,500 clients. 
This trend contrasts sharply with that in the other case studies, but the underlying reason is straightforward.
In \code{Lasp} we instrument and monitor only a subset of components, so substantial background processing occurs outside \code{ACTORCHESTRA}'s monitored path.
We confirmed this explanation by rerunning the experiments while logging heavily on the monitor side, which increased the overhead to levels comparable to those of the chat system. 
These results suggest that monitoring costs can be amortized by specifying more generic properties (rather than more fine-grained ones){; naturally, more generic properties may reduce fault-localization granularity, so the appropriate level of specificity is a trade-off that developers must balance against performance requirements.}
To further confirm this hypothesis, we added a dummy background-work function to the chat system and observed the same scaling behavior, reinforcing our conclusion.

\begin{figure}[t]
  \centering
  \includegraphics[width=\columnwidth]{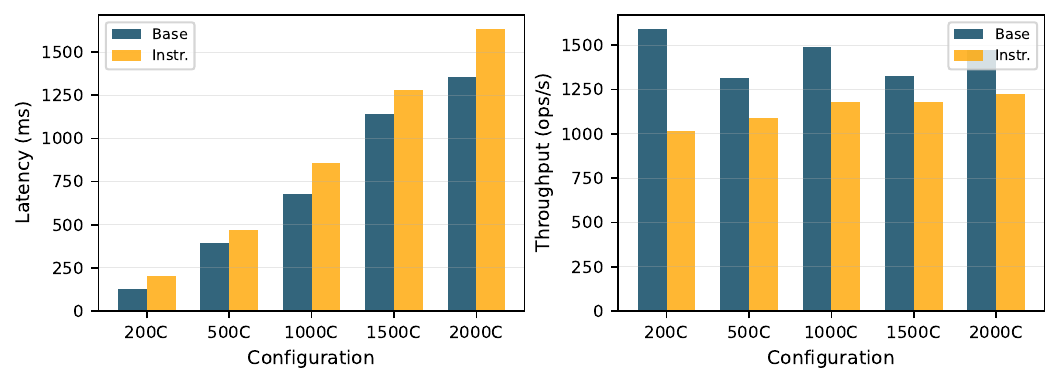}
  \caption{Lasp system latency and throughput comparison.}
  \label{fig:lasp_combined}
\end{figure}

\paragraph{Latency Analysis}
As expected, the impact on latency and throughput is less severe than in the other systems. 
Table~\ref{table:consolidated_performance} and Figure~\ref{fig:lasp_combined} list per-operation latency, showing a nearly identical pattern to execution time overhead. 
This tight correlation indicates that instrumentation adds consistent per-operation costs that scale uniformly with workload characteristics.

\paragraph{Throughput Analysis}
Table~\ref{table:consolidated_performance} and Figure~\ref{fig:lasp_combined} show throughput degradation ranging from 10.9\% to 42.4\%. 
The baseline achieves around 1,500--1,600 operations per second, and the instrumented version does not fall behind. This is significant for integration testing, correctness validation, and performance characterization during development.

\subsection{Bug Detection}

To validate \code{ACTORCHESTRA}'s detection capability, we {manually injected faults} into each system by modifying core component logic: 
\begin{enumerate*}[label=(\arabic*)]
\item the \texttt{add} callback was changed to decrement rather than increment; 
\item the chat server's room membership check was bypassed; and \item the \code{Lasp} query handler was made to return a stale counter value.
\end{enumerate*}

All violations were detected immediately upon occurrence, highlighting the key advantage of runtime verification. 
For the original correct implementations, monitors consistently reported satisfaction throughout execution, confirming no false positives across thousands of operations. We also logged all monitor behavior to a file for analysis and reporting.
%

\section{Research Question Answers and Discussion}

\textbf{RQ1}: Can causality across multi-actor interactions in OTP-compliant \code{Erlang} systems be tracked automatically? Yes, the conductor combined with compile-time AST instrumentation successfully automates causality tracking without developer intervention for OTP \code{gen\_server}-compliant systems.

While the conductor represents a potential single point of failure and bottleneck, \code{Erlang}'s supervision model and ``let it crash'' philosophy~\cite{erlangBook} provides built-in fault-tolerance mechanisms.
Leveraging these capabilities, we can deploy multiple distributed conductors to eliminate a single point of failure and alleviate bottlenecks by distributing load.

\textbf{RQ2}: Can a specification language support properties spanning multiple actor interactions while abstracting message interleavings? Yes, WALTZ enables specification of properties across multiple actors and data relations, expressing properties that span multiple interactions.

Currently, WALTZ cannot identify actors by \code{processID} separately because we use module names to guide traces. 
Therefore, properties that require variable correlations across specific process instances (e.g., ``relate this variable from client1 with this other from client2'') are not expressible. 
Supporting this would require multi-context property definitions and verification, significantly increasing semantic complexity.

We attempted to specify equivalent properties in \code{detectEr}~\cite{attard2017runtime} but encountered fundamental limitations. 
While \code{detectEr} can monitor the arithmetic system by focusing on the \texttt{add} process alone (capturing request-response pairs), thus verifying property $\varphi$, it is not possible to specify system-wide invariants in \code{detectEr} as WALTZ can. 
For example, specifying the system-wide property 
``the final result received by the client equals \code{(N + 10)*2}, where \code{N} is the original number sent by the client'' requires correlating events across the client, \texttt{main}, \texttt{add}, and \texttt{mult} modules within the same causal chain, something that is impossible in \code{detectEr}, yet possible in WALTZ. 
For the chat system and \code{Lasp}, our properties are entirely unspecifiable because \code{detectEr} can monitor only a single function within a single module, unlike \code{ACTORCHESTRA}, which monitors multiple functions. Additionally, \code{detectEr} lacks native context semantics, 
requiring developers to encode contexts manually and manage correlation tokens to avoid spurious associations between causally unrelated messages.

\textbf{RQ3}: What performance overhead is introduced by automated causality tracking and runtime monitoring? The performance characteristics must be contextualized within \code{ACTORCHESTRA}'s intended use case: development-time verification rather than production deployment. During development, engineers routinely tolerate substantial performance penalties from debugging tools in exchange for system insight. Our measured overhead of 100--145\% is reasonable for detecting causal property violations at runtime that might otherwise remain hidden until production. Recent studies on runtime overhead~\cite{overheads} confirm that the overhead of our framework is acceptable given the benefits of causality tracking.

The \code{Lasp} results show that selective instrumentation can cap overhead to 20--30\%, making \code{ACTORCHESTRA} potentially viable beyond development. 
Because overhead scales with the number of monitored components, focusing on critical paths offers a favorable trade-off between verification coverage and performance.
Our results reveal clear optimization opportunities. Throughput limitations suggest bottlenecks are addressable through decentralized conductors or distributed worker pools. Rather than general-purpose production monitoring, \code{ACTORCHESTRA} complements manual testing and debugging. In distributed systems where diagnosing subtle bugs can consume days, controlled verification overhead is worthwhile. The framework serves as a specialized runtime verification tool for development and testing, providing guarantees that conventional debugging cannot deliver.

\section{Threats To Validity}

\textbf{Internal Validity.} 
Our benchmarks were conducted on two hardware configurations (Apple M1 MacBook Air and Intel i7-10700 desktop), which may not generalize across other architectures or operating systems. 
We mitigated measurement bias through randomized execution order, 10 repetitions per configuration, and VM warm-up procedures. 
However, the deterministic workloads we designed may not capture all real-world interaction patterns. The conductor's centralized architecture inherently introduces overhead; alternative designs (e.g., distributed conductors) might yield different performance profiles but were not explored in this evaluation.

\textbf{External Validity.} Our evaluation focuses on three systems: a simple arithmetic system, a chat system, and \code{Lasp}. While these represent common actor-based patterns, they may not generalize to all distributed system architectures. Our approach targets OTP \code{gen\_server} client-server architectures specifically; systems using other OTP behaviors (e.g., \code{gen\_statem}, \code{gen\_event}) or non-OTP designs require different instrumentation strategies. Additionally, our framework assumes synchronous request-response patterns at client boundaries, which may not suit all application domains (e.g., event-driven systems, publish-subscribe architectures).

\section{Conclusion}

This work introduces \code{ACTORCHESTRA}, a runtime verification framework for actor-based systems, and WALTZ, a specification language that enables property definitions over multi-actor interactions without manual reasoning about message interleavings. By targeting \code{Erlang}'s OTP framework, the tool performs compile-time code instrumentation to propagate causality across OTP client-server systems.

Future work includes extending the tool to support additional OTP architectures (e.g., \code{gen\_statem}) and more complex communication
models, such as publish-subscribe systems, for which current causality tracking is insufficient. WALTZ will be extended to include additional language constructs (conjunctions and disjunctions), time-aware properties, and cross-context value reasoning. We aim to enhance monitor outputs with richer verdicts that explain not only that a property was violated, but also why.

\section*{Acknowledgments}
V. Mikytiv and C. Ferreira were funded by NOVA LINCS (FCT I.P. grant UID/04516/2025) and the EU Horizon TaRDIS project (grant agreement No. 101093006). B. Toninho was supported by INESC-ID (FCT I.P. grant UID/04516/2025).
\balance
\bibliographystyle{plain}
\bibliography{bibliography}

\end{document}